%% file: main.tex
\documentclass[twocolumn,aps,prl,superscriptaddress,nofootinbib]{revtex4-2}
\usepackage{graphicx}
\usepackage{color}
\graphicspath{{figures/}{fig/}}
\usepackage{amsmath}
\usepackage{amssymb}
\usepackage{bm}
\usepackage{slashed}
\usepackage{epsfig}
\usepackage{amsfonts}
\usepackage{epstopdf}
\usepackage{hyperref}
\usepackage{bbm}
\usepackage{textcomp}
\usepackage{color}
\usepackage[normalem]{ulem}

\newcommand{\sect}[1]{{\it \textbf{#1.} --- }}

\newcommand{\state}[4]{{^{#1}\hspace{-0.6mm}#2_{#3}^{[#4]}}}

\newcommand\CScSa{\state{3}{S}{1}{1}}

\newcommand\COaSz{\state{1}{S}{0}{8}}

\newcommand\COcSa{\state{3}{S}{1}{8}}
\newcommand\COcPz{\state{3}{P}{0}{8}}

\newcommand\COcPj{\state{3}{P}{J}{8}}

\begin{document}
\title{Shedding Light on Hadronization by Quarkonium Energy Correlator}

\author{An-Ping Chen}
\email{chenanping@jxnu.edu.cn}
\affiliation{College of Physics and Communication Electronics, Jiangxi Normal University, Nanchang 330022, China}

\author{Xiaohui Liu}
 \email{xiliu@bnu.edu.cn}
 \affiliation{School of Physics and Astronomy, Beijing Normal University, Beijing, 100875, China}
 \affiliation{Key Laboratory of Multi-scale Spin Physics, Ministry of Education, Beijing Normal University, Beijing 100875, China}

\author{Yan-Qing Ma}
\email{yqma@pku.edu.cn}
\affiliation{School of Physics, Peking University, Beijing 100871, China}
\affiliation{Center for High Energy Physics, Peking University, Beijing 100871, China}

\date{\today}

\begin{abstract}

We propose to measure the energy correlator in quarkonium production, which tracks the energy deposited in the calorimeter at the $\chi$-angular distance away from the identified quarkonium. The observable eliminates the need for jets while sustaining the  perturbative predictive power. Analyzing the power correction to the energy correlator, we demonstrate that the novel observable supplies a unique gateway to probing the hadronization, especially when $\cos\chi\gtrsim 0$ in the quarkonium rest frame, where the perturbative emissions are depleted due to the dead-cone effects. We expect the quarkonium energy correlator to add a new dimension to quarkonium studies.

\end{abstract}

\maketitle
\allowdisplaybreaks

\sect{Introduction}
A fundamental yet not fully understood phenomenon in quantum chromodynamics (QCD) is color confinement. This principle asserts that the colorful quarks and gluons generated in the early universe or at the collision points of particle colliders cannot be directly observed in experiments. Instead, they undergo a process called hadronization, transforming into colorless hadrons that can be detected. The question that arises is how does the hadronization process occur, or what precisely is the mechanism behind hadronization? Unfortunately, we do not yet have the answer, not even for the simplest question, how much energy is emitted during the hadronization process.

Heavy quarkonium, consisting of a heavy quark and a heavy antiquark pair ($Q\bar{Q}$), provides an ideal system for studying the hadronization mechanism. Due to their heavy mass, $Q\bar{Q}$ pairs cannot be produced or annihilated during the hadronization process. Therefore, the picture is straightforward, a $Q\bar{Q}$ pair is created in a hard collision and then undergoes hadronization to form a heavy quarkonium state. To gain insight into the specifics of hadronization, it is crucial to identify the emissions that occur during this process. To this end, some observables have been extensively studied in the literature. One approach involves examining the momentum fraction of the quarkonium production within a jet \cite{Baumgart:2014upa,Bain:2016clc,Kang:2017yde,Bain:2017wvk,Zhang:2024owr,LHCb:2017llq,LHCb:2020sey,CMS:2019ebt,CMS:2021puf,YangQian:2021}. However, this observable depends on the definition of a high transverse momentum ($p_T$) jet, which can lead to the exclusion of significant data from low $p_T$ regions. Another observable is the correlator between the quarkonium and a light hadron or hadrons~\cite{UA1:1987azc,UA1:1990eni,Porteboeuf:2010dw,ALICE:2012pet,CMS:2013jsu,CMS:2016xpm,STAR:2018smh,ALICE:2022gpu,Lansberg:2019adr}. Yet, providing a reliable theoretical prediction for this correlator can pose a significant challenge because it is not infrared-safe.

In this Letter, we propose a novel observable, the quarkonium-energy correlator. This observable is akin to the previously mentioned quarkonium-hadron correlator but is weighted by the energy of light hadrons, thereby ensuring an infrared-safe property. In the helicity frame, we demonstrate that this correlator can probe the average energy emitted during the hadronization process and distinguish between different production mechanisms, like the color-singlet (CS) or the color-octet (CO) mechanisms in the Non-Relativistic QCD (NRQCD)~\cite{Bodwin:1994jh,Beneke:1997av,Brambilla:1999xf,Kramer:2001hh,Brambilla:2004jw,Brambilla:2010cs}.

\begin{figure}[htbp]
 \begin{center}
 \vspace*{0.cm}
 \hspace*{-5mm}
 \includegraphics[width=0.25\textwidth]{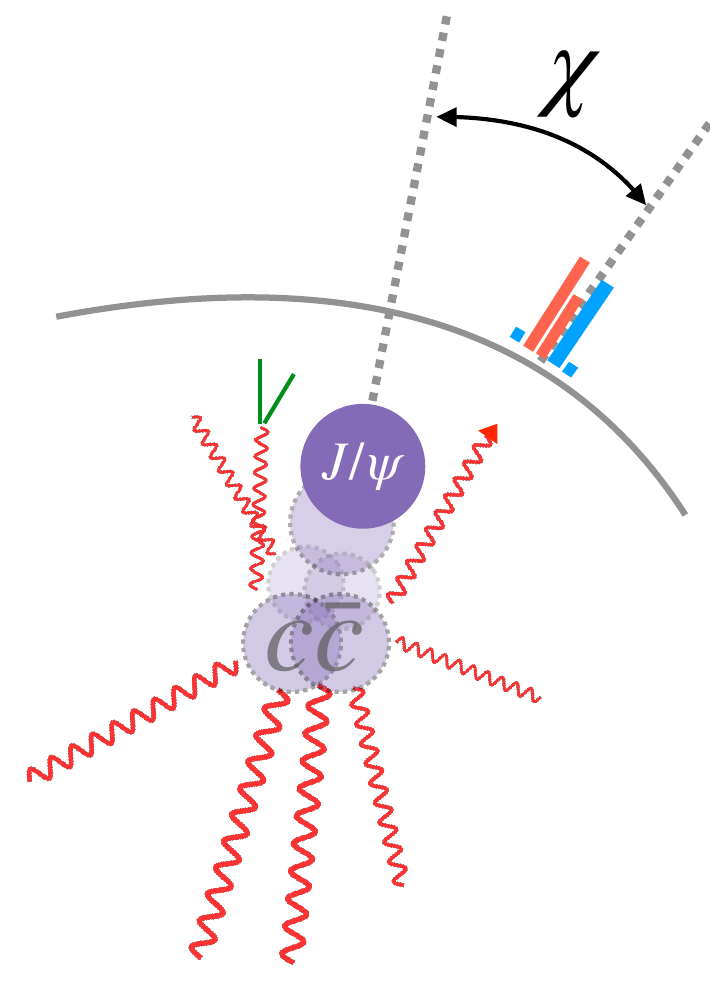}
 \end{center}
 \vspace*{-.5cm}
 \caption{Measurement of the energy correlator around the $J/\psi$, where the energy deposit in the calorimeter with the polar angle separation $\chi$ from the $J/\psi$ is recorded. $\chi$ is measured in the $J/\psi$ rest frame.  \label{fig:jpsi-ec}}
 \vspace*{0.cm}
\end{figure}

\begin{figure}[htb!]
 \begin{center}
 \vspace*{0.cm}
 \hspace*{-5mm}
 \includegraphics[width=0.45\textwidth]{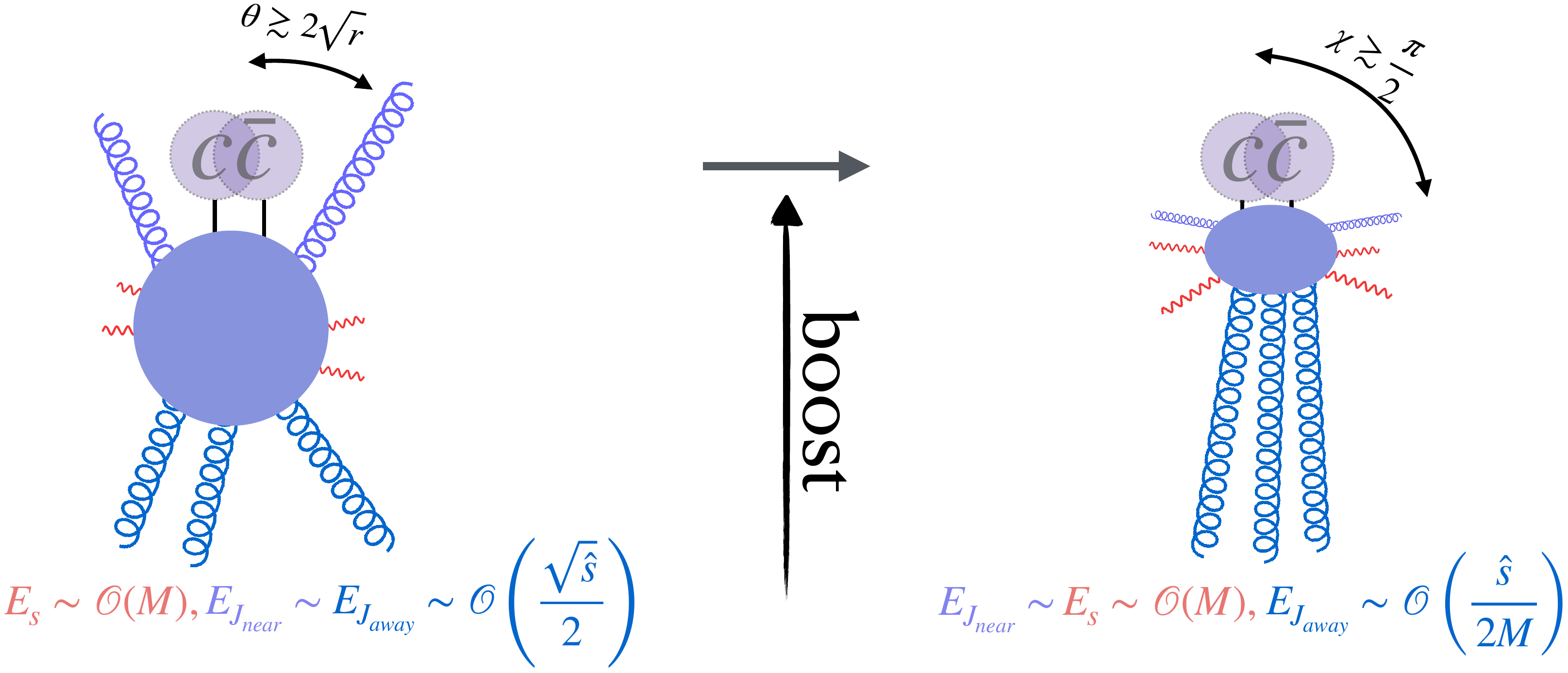}
 \end{center}
 \vspace*{-.5cm}
 \caption{ The dominant configuration for hard emissions when $\sqrt{\hat{s}} \gg M$ in the center of mass frame (left) and the quarkonium rest frame (right). \label{fig:jpsi-boost}}
 \vspace*{0.cm}
\end{figure}

\sect{Quarkonium energy correlator}
We consider a generic quarkonium ${\cal Q}$ (e.g., $J/\psi$) production process, such as in $e^+e^-$ annihilation, hadronic collisions, or $B$ meson decay. In addition to the quarkonium detection,
we measure the total energy flow along a specific angular direction $\chi$, relative to the flying direction of the ${\cal Q}$ but boosted into its rest frame, the so-called helicity frame, see Fig.~\ref{fig:jpsi-ec} and Fig.~\ref{fig:jpsi-boost} for illustration. Our focus lies on measuring the {\it Quarkonium Energy Correlator} expressed as,
\begin{align}\label{eq:eec-def}
\Sigma(\cos\chi) = \int d\sigma \,\sum_i \frac{E_i}{M} \delta(\cos\chi - \cos\theta_i)\,,
\end{align}
where $d \sigma$ is the differential cross section responsible for generating the ${\cal Q}$, weighted by the total energy $\sum_i E_i \delta(\cos\chi-\cos\theta_i)$ carried by particles propagating at
the angle $\chi$ with respect to the ${\cal Q}$. Here, the sum over $i$ could include either all final state hadrons or its pre-selected subset by a fat jet.
We normalize this energy to $M$, the mass of the quarkonium. The measurement process for the $J/\psi$ energy correlator is outlined in Fig.~\ref{fig:jpsi-ec}. Eq.~(\ref{eq:eec-def}) implies that the energy correlator can be understood as the average energy radiation flow directed towards the given angle $\chi$. Therefore, by looking at the detectors placed at various $\chi$ positions, we can explore the radiation pattern in ${\cal Q}$ production through $\Sigma(\cos\chi)$.

In principle, $\Sigma(\cos\chi)$ receives contributions from both hard radiation during the collision and non-perturbative soft radiation during the $Q{\bar Q}$ pair's transition into the ${\cal Q}$, in the form of~\cite{Korchemsky:1999kt,Belitsky:2001ij,Schindler:2023cww}
\begin{align}\label{eq:NPform}
\Sigma(\cos\chi) =
\Sigma_{P.T.}(\cos\chi)
+ \Sigma_{N.P.}(\cos\chi) \,.
\end{align}
Theoretically, the energy weighting ensures the infrared safety of the observable, enabling the calculability of the hard radiation contribution $\Sigma_{P.T.}$ using perturbation theory. As we will show, $\Sigma_{P.T.}$ is typically dominant when $\chi \gtrsim \frac{\pi}{2}$ and can serve as a benchmark to verify the theoretical framework and the accuracy of experimental data. Meanwhile, the non-perturbative soft radiation contribution $\Sigma_{N.P.}$ contains the physics necessary for comprehending the hadronization mechanism.
As we will demonstrate in the rest of this Letter, the non-perturbative contribution $\Sigma_{N.P.}$ could be significant when $\chi \lesssim \frac{\pi}{2}$ compared with $\Sigma_{P.T.}$.

To estimate the significance of $\Sigma_{N.P.}$, we will first examine the scaling behavior of radiation illustrated in Fig.~\ref{fig:jpsi-boost}. Starting with the center of mass frame generating the ${\cal Q}+X$ (sub)-system, with an input energy $\sqrt{\hat{s}} \gtrsim M$, the production configuration is dominated by $2$ back-to-back jets with energy $\sim {\cal O}(\frac{{\sqrt{\hat{s}}}}{2})$ and transverse broadening $\sim{\cal O}(M)$. The inter-jet region is filled with soft radiation with momentum $\sim {\cal O}(M)$, due to the soft and collinear feature of perturbative QCD. The collinear radiation near the quarkonium is predominantly confined to the region $\sin \theta \gtrsim \frac{M}{\sqrt{\hat{s}}/2} \equiv 2\sqrt{r} $ due to the dead-cone effects~\cite{Dokshitzer:1991fd,ALICE:2021aqk}. When boosted into the quarkonium rest frame, the momentum transverse to the jet axis remains unchanged. The energy of the backward jet gets magnified due to a large boost factor $a \sim \frac{\sqrt{\hat{s}}}{M} = \sqrt{\frac{1}{r}}$ to reach $E_{J_{away}} \sim a \frac{\sqrt{\hat{s}}}{2}  = \frac{\hat{s}}{2M}$. The energy for the collinear radiation near the quarkonium is reduced to $E_{J_{near}} \sim {\cal O}(M)$, along with the quarkonium, and the radiation is pushed to the region $\chi \gtrsim {\cal O}\left(\frac{\pi}{2}\right)$. We thus expect when moving from $\cos \chi \sim -1$ to $\cos\chi \sim 0$, the magnitude of $\Sigma_{P.T.}$ drops rapidly by a factor of $\sim {\cal O}\left(E_{J_{away}}/E_{J_{near}}\right) = {\cal O}\left(\frac{1}{2r} \right)$, and also gets suppressed as we approach $\cos\chi \sim 1$ due to the presence of the dead-cone.

The depopulation of perturbative radiation in the $\cos\chi >0$ region allows $\Sigma_{N.P.}$ to potentially outperform $\Sigma_{P.T.}$. Therefore, we focus on this region and analyze the relative importance between  $\Sigma_{N.P.}$  and $\Sigma_{P.T.}$.

Firstly, we note that one can intuitively understand the additive form in Eq.~(\ref{eq:NPform}) by considering that $\Sigma(\chi) \sim \int^M E_i \, d\sigma \sim \left(\int^M_{\Lambda} + \int^\Lambda\right) E_i \, d\sigma $, where $\Lambda$ separates the perturbative hard ($E_i > \Lambda$) radiation from the non-perturbative soft ($E_i < \Lambda$) radiation. This implies that the size of $\Sigma_{N.P.}$ can be approximated by replacing the corresponding hard emission $i$ with a non-perturbative soft emission. For instance, a hard emission associated with a power of $\alpha_s(M)$ contributes to
$\Sigma_{P.T.}$ on the order of $ \alpha_s(M) \frac{M} {M} M^2$, where $M^2$ estimates the hard emission phase space volume. By substituting the hard emission with a soft one, we can estimate $\Sigma_{N.P.} \sim \alpha_s(\Lambda) \frac{\Lambda}{M} \Lambda^2 $ where $\Lambda^2$ approximates the soft radiation phase space volume. The relative size is then $\Sigma_{N.P.}/\Sigma_{P.T.} \sim  \frac{\alpha_s(\Lambda)}{\alpha_s(M)}\frac{\Lambda^3}{M^3}$.

To quantify the relative size, we estimate it within the NRQCD framework, where
different production channels contribute to $\Sigma(\cos\chi)$,
\begin{align}\label{eq:nonpt}
\Sigma(\cos\chi) = \sum_{ch}
\Sigma_{P.T.}^{ch}(\cos\chi)
+ \Sigma_{N.P.}^{ch}(\cos\chi) \,,
\end{align}
with $ch$ running over the CS and CO channels, proportional to their long-distance matrix elements (LDMEs), respectively.
When replacing a hard emission with a soft one, a transition between channels can occur since the soft contribution is integral to the $Q{\bar Q}$ LDMEs in NRQCD.
We focus on the CS contributions to $\Sigma_{P.T.}$, and after the replacement one switches the CS channel to a CO one. Since CO LDMEs are typically smaller than CS LDMEs~\cite{Bodwin:1994jh,Bodwin:2012xc}, this usually results in a smaller estimation of $\Sigma_{N.P.}/\Sigma_{P.T.}$. Following the previous example, the related power counting is $\Sigma_{P.T.} \sim \alpha_s(M)M^2 \langle  {\cal O}_1 \rangle $ and $\Sigma_{N.P.} \sim v \times \alpha_s(Mv)  (Mv)^2 \langle  {\cal O}_1 \rangle  \sim v M^2 \langle {\cal O}_8 \rangle $, where we have set $\Lambda = Mv$ and $\langle {\cal O}_{1(8)} \rangle$ denotes the CS (CO) LDMEs. Also, $ \langle {\cal O}_8 \rangle \sim \alpha_s(Mv) v^2 \langle {\cal O}_1 \rangle $~\cite{Bodwin:1994jh,Bodwin:2012xc}, and $v \ll 1$ is the typical relative velocity between the $Q\bar{Q}$ pair of the quarkonium. For a charmonium system, $v^2 \sim 0.3$ and $\alpha_s(M) \ll \alpha_s(Mv)\sim v$~\cite{Bodwin:1994jh}, therefore $\Sigma_{N.P.}/\Sigma_{P.T.} \sim
 \frac{v^4}{\alpha_s(M)} \sim 0.4$, making $\Sigma_{N.P.}$ sizable. The relative size of $\Sigma_{N.P.}$ can be further enhanced due to the dead-cone effects of perturbative hard radiation~\cite{Dokshitzer:1991fd,ALICE:2021aqk}. It is then promising to detect the hadronization effects.

Notably, higher-order corrections in $\alpha_s$ or $v$ should be modest for $\Sigma_{N.P.}/\Sigma_{P.T.}$, as they influence the $d\sigma$ in $\Sigma(\chi) \sim \left(\int^M_{Mv} + \int^{Mv}\right) E_i \, d\sigma$, whose effect largely cancels out in the ratio.

 Therefore, $\cos\chi > 0$ is a region where the hadronization effect can be significant, potentially rivaling the contributions from hard emissions. This suggests that detecting energy emitted during the hadronization is feasible by measuring the quarkonium energy correlator in the domain where $\cos\chi > 0$. To further illustrate this concept, the following sections will provide explicit examples using $J/\psi$ production.

\sect{Theoretical predictions for quarkonium energy correlators}
We first exemplify our proposal using the $J/\psi$ production in $e^+e^-$ annihilation, $e^+e^- \to \gamma^\ast \to J/\psi + X$. To provide a theoretical prediction, we employ the NRQCD factorization framework~\cite{Bodwin:1994jh}, although other methods can also be used \cite{Ellis:1976fj,Carlson:1976cd,Chang:1979nn,Fritzsch:1977ay,Halzen:1977rs,Ma:2016exq,Ma:2017xno}. For this problem, the most important contributions come from $^3S_1^{[1]}$, $^1S_0^{[8]}$ and $^3P_J^{[8]}$ channels.

The leading hard process for the CS channel $^3S_1^{[1]}$ is $e^+e^-\to \gamma^\ast \to c{\bar c}[^3S_1^{[1]}] +g+g$, and thus the hard emissions can contribute to the $\cos\chi>0$ region. The leading hard process for the CO channel $^1S_0^{[8]}$ or $^3P_J^{[8]}$ is $e^+e^-\to \gamma^\ast \to c {\bar c}[^1S_0^{[8]},^3P_J^{[8]}] +g$, where the hard gluon and the $c\bar{c}$ pair are back-to-back, and the hard emission only contributes to $\cos\chi=-1$. Furthermore, soft emissions during hadronization of the CS $c{\bar c}$ pair are generally expected to be less than those from the CO $c{\bar c}$ pairs. Therefore, the leading $\Sigma_{P.T.}$ comes from the $^3S_1^{[1]}$ channel, and  the leading $\Sigma_{N.P.}$ comes from the $^1S_0^{[8]}$ and $^3P_J^{[8]}$ channels. Based on the previous general discussion, we expect that the  $\Sigma_{N.P.}$ can be significant, as will be shown by explicit computation.

For the leading contribution to $\Sigma_{P.T.}$, Eq.~(\ref{eq:eec-def})  becomes
\begin{align}\label{eq:eePT}
\Sigma_{P.T.}^{^3S_1^{[1]}} (\cos\chi)
&= \int d\hat{\sigma}_{e^+e^-\to c{\bar c}[^3S_1^{[1]}]+g+g}
\langle {\cal O}^{J/\psi}(^3S_1^{[1]})
\rangle  \nonumber \\
& \hspace{5.ex}
\times \sum_{i=1}^2 \frac{E_i}{2m_c}\delta(\cos\chi - \cos\theta_i)\,,
\end{align}
where $d\hat{\sigma}$ is the leading order (LO) partonic cross section that generates the CS $c{\bar c}$ pair,
$\langle {\cal O}^{J/\psi}(^3S_1^{[1]})
\rangle$ is the NRQCD matrix element, and we have replaced the quarkonium mass $M$ with $2m_c$ in the fixed order calculation by neglecting the binding energy \cite{Bodwin:1994jh}. $\theta_i$ is the angle between the directions of one of the two gluons and the $J/\psi$. The analytic results can be found in the Supplemental Material (SM).

To provide an estimate for $\Sigma_{N.P.}$, we need to model the hadronization process from a CO $c\bar{c}$ pair to $J/\psi$. In NRQCD, it is suggested that the transition is dominated by one soft gluon emission, while multiple emissions are subleading in $v$. If we denote the transition amplitude for $c{\bar c}[ch] \to J/\psi + g$ as $A^{ch}(\theta,\phi)$ with $ch = \COaSz, \COcPj$, then we find
\begin{align}\label{eq:eec-had}
\Sigma^{ch}_{N.P.}(\cos\chi) =& \int {d \Phi_g }\frac{k^0}{M}
\delta(\cos{\chi} - \cos{\theta})
\sum_{\lambda}
\hat{\sigma}^{ch}_{\lambda} \,
A^{ch}_\lambda(\theta,\phi),
\end{align}
where $\Phi_g$ denotes the gluon phase space, $k^0$ represents the emitted energy, and $\hat{\sigma}^{ch}$ is the total cross section that produces the $c{\bar c}[ch]$ pair with polarization $\lambda$. We can see that
$\int \frac{k^0 d\phi}{16\pi^3}
 \sum_{ch, \lambda}
\frac{\hat{\sigma}^{ch}_{\lambda}}{\sigma} \,
A^{ch}_\lambda(\theta,\phi)
 $ characterizes the radiated energy distribution $\rho(k_0,\theta)$ due to hadronization.

As for the $\COaSz$ channel, since the intermediate $c\bar c[\COaSz]$ state is unpolarized, the soft gluon emission should be isotropic, therefore we have
$A^{\COaSz}(\theta,\phi) = \frac{1}{4\pi}G_0$.
Here $G_0$ is a function of $k^0$ but  independent of $\theta$ and $\phi$. $G_0$ describes the probability of the $c{\bar c}$ pair emitting a gluon to form the $J/\psi$, therefore one should have $\frac{1}{4\pi}\int d\Phi_g G_0 \sim {\cal O}(\langle {\cal O}^{J/\psi}(\COaSz) \rangle) $. Hence, we find the hadronization contribution to the energy correlator to be isotropic with the size
\begin{align}\label{eq:ee1s0}
\Sigma^{\COaSz}_{N.P.}(\cos\chi)
= &
{\hat{\sigma}^{\COaSz}}
\int
d\Phi_g \frac{k_0}{M}
\frac{G_0}{4\pi} { \delta(\cos{\chi} - \cos{\theta})}
\nonumber\\
   \equiv &  {\bar v}_0  \frac{\sigma^{\COaSz}}{ 4  }  \,,
\end{align}
where $m_c {\bar v}_0$ parameterizes the average energy emitted during hadronization. $\hat{\sigma}^{\COaSz}$ and $\sigma^{\COaSz}$ are the total cross sections for producing the $c{\bar c}[\COaSz]$ pair and the $J/\psi$ through that $c{\bar c}$ channel, respectively. At LO, we have
 $ \sigma^{\COaSz}
=
\frac{32 \pi ^2 \alpha^2 \alpha _s  e_c^2  }{3
	s^2 m_c}(1-r) \langle {\cal O}^{J/\psi}(\COaSz) \rangle$.
The contribution from the $\COcPj$ states can be obtained similarly. The calculation is more involved and we leave the details in the SM.

\begin{figure}[htb!]
 \begin{center}
 \vspace*{0.cm}
 \hspace*{-5mm}
 \includegraphics[width=0.5\textwidth]{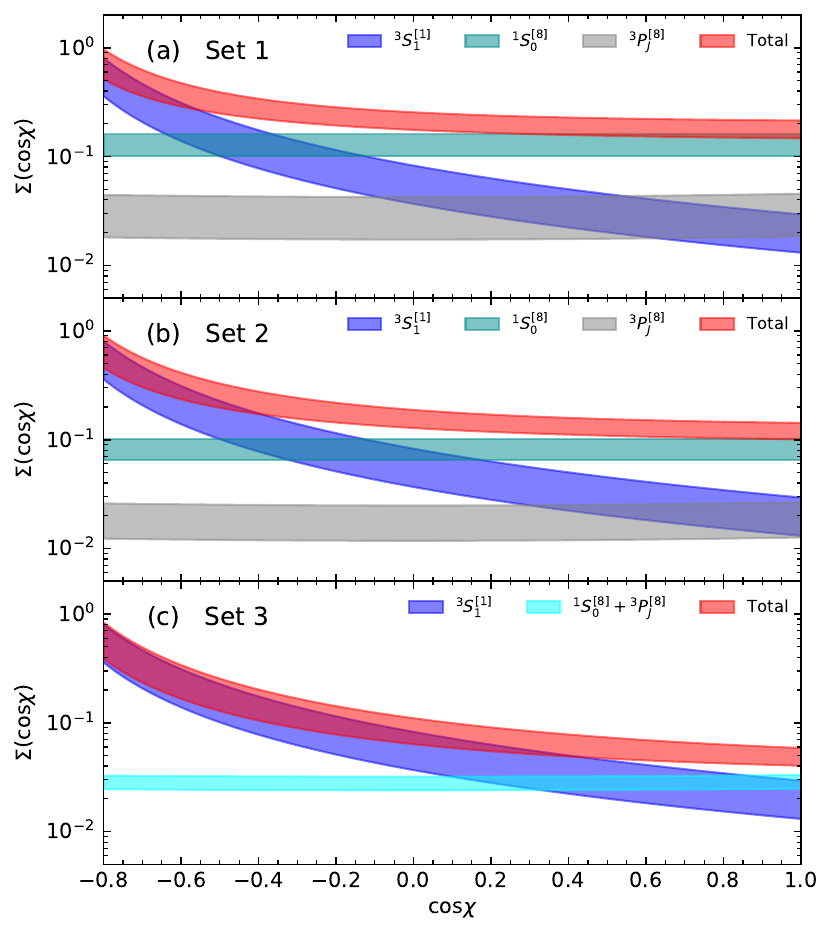}
 \end{center}
 \vspace*{-.5cm}
 \caption{ Theoretical predictions for $J/\psi$ energy correlator in $e^+e^-$ collision with $\sqrt{s}=10.6$ GeV. Here, $\chi$ is defined in the $J/\psi$ rest frame. \label{fig:ee}}
 \vspace*{0.cm}
\end{figure}

Fig.~\ref{fig:ee} exhibits the numerical consequence of our analysis. We set $\sqrt{s} = 10.6 \, {\rm GeV}$ to match the Belle~\cite{Belle:2009bxr} kinematics and $\mu=\sqrt{s}/2$ to evaluate $\alpha_s$. We use $\langle  \mathcal{O}^{J/\psi}(\CScSa)  \rangle=1.16 \, \mathrm{GeV}^3$ calculated from the potential model~\cite{Eichten:1995ch}. For the CO LDMEs, we choose two sets of LDMEs extracted from hadron colliders~\cite{Chao:2012iv,Feng:2018ukp}, as shown in Set 1 and 2 in Table.~\ref{table:LDMEs}.
\begin{table}[hbt]
\caption{ NRQCD LDMEs used in this Letter.}
\label{table:LDMEs}
\begin{center}
  \begin{tabular}{  l | c | c | c }
  \hline
     & $\langle {\mathcal O}^{J/\psi}(^1S_0^{[8]})\rangle$ & $\langle {\mathcal O}^{J/\psi}(^3S_1^{[8]})\rangle$ & $\langle {\mathcal O}^{J/\psi}(^3P_0^{[8]})\rangle/m_c^2$ \\
     & $10^{-2}$ GeV$^3$ & $10^{-2}$ GeV$^3$ & $10^{-2}$ GeV$^3$
        \\ \hline
    Set 1\cite{Chao:2012iv}  & $8.9\pm0.98$ & $0.30\pm0.12$  & $0.56\pm0.21$ \\ \hline
    Set 2\cite{Feng:2018ukp} & $5.66\pm0.47$ & $0.177\pm0.058$ & $0.342\pm0.102$\\ \hline
    Set 3 & $1.0$ &  & $0.25$\\ \hline
  \end{tabular}
\end{center}
\end{table}
There exist other fits to the LDMEs~ \cite{Butenschoen:2011yh,Gong:2012ug,Bodwin:2015iua,Brambilla:2022rjd}, but the fitted values are not all positive, therefore we do not include them for comparison here.

To evaluate $\Sigma_{N.P.}$, we tentatively set $\bar{v}_0^2 =\bar{v}_1^2 =0.25$ consistent with NRQCD power counting for charmonia, while its true value should be determined by future experimental analysis. Besides the
errors from the LDMEs, we also include the scale uncertainty by varying $\mu$ up and down by a factor of $2$.

The blue bands in Fig.~\ref{fig:ee} show the LO $^3S_1^{[1]}$ contribution to the $\Sigma(\cos\chi)$.
We can see that the radiation in the CS contribution occupies dominantly around $\chi \sim \pi$ and drop significantly when $\chi \lesssim \frac{\pi}{2}$, which coincides with our previous analysis.
The suppression of the hard radiation increases the likelihood of probing the non-perturbative hadronization contribution when $\chi\lesssim \frac{\pi}{2}$. As manifested in Fig.~\ref{fig:ee}, $\Sigma_{N.P.}$ is significant for all choices of LDMEs, indicating the detectability of the hadronization effect with the current Belle experiment \cite{Belle-II:2018jsg}.
We note the unnaturally large $\Sigma_{N.P.}$ from the $\COaSz$ channel in Sets 1 and 2 is due to
the known universality problem in NRQCD~\cite{Ma:2017xno} that
the values of $\langle {\mathcal O}^{J/\psi}(^1S_0^{[8]})\rangle$ extracted from hadron colliders are significantly larger than the upper bound set by $J/\psi$ production in
$e^+e^-$ collisions~\cite{Zhang:2009ym}.
To provide a more reasonable prediction for $e^+e^-$ collision, we also include a set of CO LDMEs that satisfies the upper bound set by the $B$ factories \cite{Zhang:2009ym}, denoted as Set $3$ in Table.~\ref{table:LDMEs}.
With this Set, the $\Sigma_{N.P.}$ is still significant.

\begin{figure}[htb!]
 \begin{center}
 \vspace*{0.0cm}
 \hspace*{-5mm}
 \includegraphics[width=0.5\textwidth]{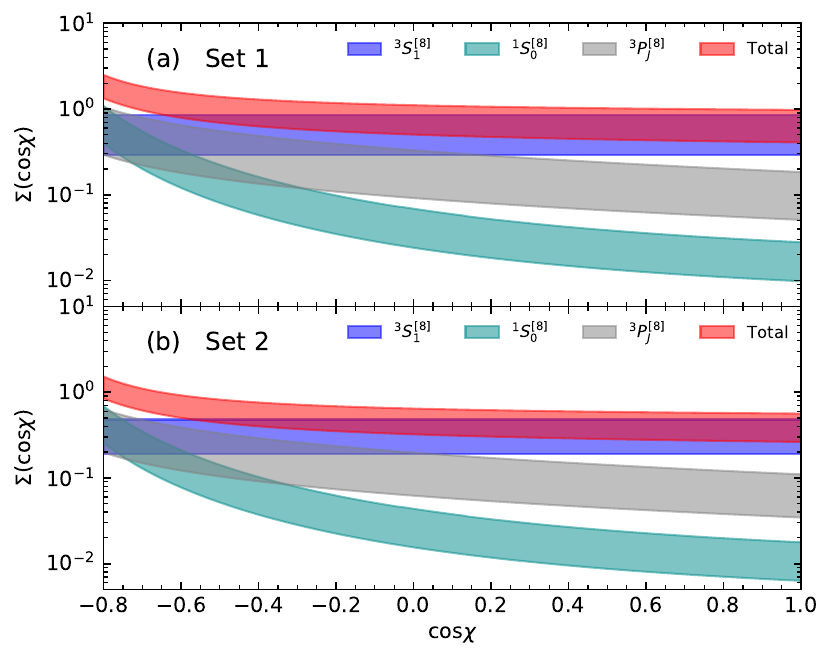}
 \end{center}
 \vspace*{-.5cm}
 \caption{ Theoretical predictions for $J/\psi$ energy correlator in $pp$ collision.
 Contribution from
the short-distance coefficient $\hat{\sigma}_{A+B\to g+X}$ is not included in this prediction. \label{fig:pp}}
 \vspace*{-.1cm}
\end{figure}

Now we show the energy correlation in $pp$ collisions. For illustration, we consider $J/\psi$ production at high $p_T$ and  simplify our analysis using a single-gluon fragmentation framework, in which
\begin{align}\label{eq:pp-m}
\Sigma(\cos{\chi}) =& \sum_{n} \int_0^1 dz
d{\hat{\sigma}}_{A+B\to g+X}({\hat p}/z,\mu_F) \nonumber\\&\times
 \hat D_{g\to c\bar c[n]}(z,\cos{\chi},\mu_F)
\langle  \mathcal{O}^{J/\psi}(n)  \rangle\,.
\end{align}
The notations and the derivation can be found in the SM. { Since we are mostly interested in the region $\cos\chi>0$, for illustrative purposes, we have ignored the contribution to $\Sigma(\cos\chi)$ from the short coefficient $\hat{\sigma}_{A+B\to g+X}$ which typically contributes to $\cos\chi <0$. A complete analysis will be presented in future publications.} The numerical results are exhibited in Fig.~\ref{fig:pp}, obtained with $p_T=15\, \textrm{GeV}$ and $\vert y \vert<0.9$  at $\sqrt{s}=7\, \textrm{TeV}$. Here $\Sigma_{P.T.}$ receives dominant contributions from the $^1S_0^{[8]}$ and $^3P_J^{[8]}$ channels, and $\Sigma_{N.P.}$ from the $^3S_1^{[8]}$ channel. We have set $\bar v_2^2=0.25$ and the factorization scale $\mu_f=\mu = p_T$. Besides the
errors from the LDMEs, we also include the scale uncertainty by varying $\mu$ and $\mu_f$ up and down by a factor $2$.
The calculated $\Sigma_{P.T.}$ supports our previous analysis based on the generic properties of QCD. A significant $\Sigma_{N.P.}$ is observed from the $\COcSa$ channel in our current predictions.
This is because the partonic cross-section $d\hat\sigma_{A+B\to g+X} \sim z^4$ \cite{Mangano:1996kg,Bodwin:2015iua}. Consequently, the energy correlator effectively scales as the 4th moment of the fragmentation functions (FFs) $\hat{D}_{g\to c{\bar c}[n]}$. Compared to the LO NRQCD $\COcSa$ FF which is proportional to $\delta(1-z)$, the $\COaSz$ FF peaks at $z=0$ and the $\COcPj$ FF gradually increases as $z\to 1$. Therefore when convoluted with $d\hat\sigma_{A+B\to g+X}$, the contributions from $\COaSz$ and $\COcPj$ are significantly suppressed, compared with $\COcSa$. Relativistic corrections at higher orders will soften and migrate the peak in the $\COcSa$ FF to lower values of $z$ and hence reduce the estimated $\Sigma_{N.P.}$ size. For a more reliable estimation, one can apply the soft gluon factorization~\cite{Ma:2017xno} to account for higher order relativistic corrections. By doing so, one finds the hadronization effect to be reduced by a factor of $60\%$, but the size remains detectable for future measurements.

To close the section, we note that our prediction assumes the rotational covariance of the non-perturbative radiation as required by the NRQCD factorization theorem. Although this assumption has been explicitly verified up to 2-loop order~\cite{Ma:2005cj,Nayak:2006fm,ftnt}, the rotational covariance of the NRQCD soft emission necessitates further testing and verification. The quarkonium energy correlator provides the possibility to probe the distribution of the soft radiation experimentally. For instance, instead of rotational covariance, if the soft radiation is boost-covariant, which is the case where the soft radiation relies explicitly on the direction of a Wilson line~\cite{Nayak:2005rw}, we may expect $\Sigma_{N.P.}  \sim \frac{1}{\sin^3\chi} $ similar to light hadron energy-energy correlator~\cite{Korchemsky:1999kt,Belitsky:2001ij,Schindler:2023cww} and experiences significant amplification as $\chi \to 0$, which is distinct from the expectation based on the rotational covariance assumption. As a result, future experiments could differentiate between these distributions through a refined measurement near $\chi = 0$ in the quarkonium energy correlator spectrum.

\sect{Summary and outlook}
In this Letter, we propose the
quarkonium-energy correlator, which extends the concept of energy correlators~\cite{Basham:1978bw,Basham:1978zq,Hofman:2008ar,Belitsky:2013ofa,Belitsky:2013xxa,Kologlu:2019mfz,Korchemsky:2019nzm,Dixon:2019uzg,Chen:2019bpb,Chicherin:2020azt,Chen:2020adz,Chen:2020vvp,Chang:2020qpj,Li:2021zcf,Jaarsma:2022kdd,Komiske:2022enw,Holguin:2022epo,Yan:2022cye,Chen:2022jhb,Chang:2022ryc,Chen:2022swd,Lee:2022ige,Liu:2022wop,Ricci:2022htc,Yang:2022tgm,Andres:2022ovj,Craft:2022kdo,Chen:2022pdu,Liu:2023aqb,Li:2023gkh,Andres:2023xwr,Devereaux:2023vjz,Andres:2023ymw,Jaarsma:2023ell,Lee:2023npz,Lee:2023tkr,Yang:2023dwc,Chen:2023zzh,Holguin:2023bjf,Cao:2023qat,Gao:2023ivm,Barata:2023zqg,Barata:2023bhh,Barata:2024nqo,Yang:2024gcn,Liu:2024kqt,Csaki:2024joe} to heavy quarkonium studies. The quarkonium energy correlator measures the energy flow at an angle distance $\chi$ from the quarkonium in its rest frame. Notably, for generic $J/\psi$ production, such as at the Belle experiment or the LHC, we demonstrate that the hadronization effect leads to a power correction to the energy correlator comparable to that of the perturbative hard radiation in the region where $\cos\chi \gtrsim 0$, due to the compelling dead-cone effects. Our explicit calculations support this claim, further demonstrating its potential to unveil the hadronization mechanisms in heavy quarkonium production. We thus expect the energy correlator to be an exceptional observable to offer unprecedented insights into the complex non-perturbative phenomena governing heavy quarkonium production.

\begin{acknowledgments}
\sect{Acknowledgments}
    The work was supported in part by the National Natural Science Foundation of
	China (No. 12325503, No. 12205124), the National Key Research and Development Program of China under
	Contract No. 2020YFA0406400, and the High-performance Computing Platform of Peking University. X.~L. is supported by the Natural Science Foundation of China under Contract No.~12175016.
\end{acknowledgments}

\input{paper.bbl}

\newpage
\newpage
\newpage
\appendix
\begin{widetext}
\section{Supplemental Materials for ``Shedding Light on Hadronization by Quarkonium Energy Correlator"}

\subsection{Quarkonium Energy Correlator in $e^+e^-$ annihilation}
We present analytic results for the energy correlator measured in the $J/\psi$ leptonic production.

Following Eq.~(\ref{eq:eePT}), we
boost $J/\psi$ to its rest frame, and define
\begin{align}
\cos\theta_i
= \frac{1}{
\sqrt{z^2 -4 r }}
\left(\frac{2  x_i }{z_i}
- z \right)
\,,
\end{align}
where $r = \frac{4m_c^2}{s}$ and $\sqrt{s}$ represents the machine center of mass energy. We have introduced the dimensionless variables $z = \frac{2p\cdot q}{s}$,
$x_i = \frac{2k_i\cdot q}{s}$, and $z_i = \frac{2k_i\cdot p}{4m_c^2}$, with $q$, $p$  and $k_i$ denoting the momentum of the virtual photon, the $c{\bar c}$ pair  and the out-going gluons, respectively.

The phase space integration in Eq.~(\ref{eq:eePT}) can be parameterized in terms of $x_1$ and $z$, which gives
\begin{align}\label{eq:sig1}
& \Sigma_{P.T.}^{^3S_1^{[1]}} (\cos\chi)
=  \frac{256 \pi e_c^2  \alpha^2 \alpha_s^2}{ 81  s }
\frac{{\cal O}^{J/\psi}(^3S_1^{[1]})}{(2m_c)^3} r^2
 \nonumber \\
&
\times
 \int_{2\sqrt{r}}^{1+r}
 dz \int_{x_{-}}^{x_+}\, dx_1  f(z,x_1)
 \sum_{i=1}^2  \frac{z_i}{2}   \delta(\cos\chi - \cos\theta_i)\,,
\end{align}
where we have replaced $E_i/(2m_c)$ in the $J/\psi$ rest frame with $z_i/2$. Here, $z_1 = \frac{1}{2r} (x_1-x_2+z-2r)$, $z_2 = \frac{1}{2r} (x_2-x_1+z-2r)$, $x_2 = 2-x_1-z$, $x_\pm = \frac{1}{2} \left(2-z\pm \sqrt{z^2-4r}\right)$ and
\begin{align}
&f(z,x_1)=  \frac{1}{(2-z)^2}  \Big( \frac{1}{r}+
\frac{(2+x_1)x_1}{ (1-x_2-r)^2} +
\frac{(2+x_2)x_2}{ (1-x_1-r)^2} \nonumber \\
& + \frac{2(1-z)(1-r)}{r(1-x_2-r)(1-x_1-r)}
+  \frac{(2-z)^2((z-r)^2-1)}{(1-x_2-r)^2(1-x_1-r)^2} \nonumber \\
& +
 \frac{6(1+r-z)^2}{(1-x_2-r)^2(1-x_1-r)^2} \Big)     \,,
\end{align}
is known in the literature~\cite{Keung:1980ev,Yuan:1996ep}. Evaluating Eq.~(\ref{eq:sig1}) finds the LO color singlet contribution to the energy correlator.

The non-perturbative $\Sigma_{N.P.}$ also receives a contribution from the $\COcPj$ states where,
 at the leading order in $v$, a $c\bar c[\COcPj]$ pair with $S=1$ and $L=1$ transmits into
the $J/\psi$ through one electrical dipole ($E_1$) interaction. We note that different helicity and angular momentum states contribute differently to the polar $\chi$ angle distribution. Following~\cite{Cho:1994gb} and using the heavy quark spin symmetry, we find
\begin{subequations} \label{eq:3pj8}
\begin{align}
A_{T}^{\COcPj}(\theta,\phi)=&  \frac{3(1+\cos^2\theta)}{16\pi} G_1
,\\
A_{L}^{\COcPj}(\theta,\phi)=&  \frac{3(1-\cos^2\theta)}{8\pi} G_1\,,
\end{align}
\end{subequations}
where the subscripts $T$ and $L$ stand for the $z$-component of the $c{\bar c}$-pair's orbital angular momentum $|L_z|=1$ and $0$, respectively.
$G_1$ is solid angle independent and satisfies $\frac{1}{4\pi}\int d\Phi_g G_1 \sim {\cal O}(\langle {\cal O}^{J/\psi}(\COcPz) \rangle) $. We then derived
the hadronization contribution in the $\COcPj$ channel to be
 \begin{align}\label{eq:ee2}
\Sigma_{N.P.}^{\COcPj} (\cos{\chi}) =&
\int d\Phi_g \frac{k^0}{M} \frac{G_1}{ 2\pi}   \delta(\cos{\chi} - \cos{\theta})
\times 3 \left(\frac{1+\cos^2\theta}{8}{\hat {\sigma}_{T}^{\COcPz}}
  +\frac{1-\cos^2\theta}{4} {\hat{\sigma}_{L}^{\COcPz}} \right) \nonumber \\
  & \hspace{-2.5cm}\equiv  {\bar v}_1
  \times
  3 \left(\frac{1+\cos^2\chi}{16}{  {\sigma}_{T}^{\COcPz}}
  +\frac{1-\cos^2\chi}{8} { {\sigma}_{L}^{\COcPz}}  \right) \,.
\end{align}
Here $m_c \bar{v}_1$ is the average energy emitted and the $J/\psi$ total production cross section is given by
\begin{subequations}
\begin{align}
 {\sigma}_{T}^{\COcPz}
=&
-F
 \frac{2  }{(1-r)^4}
    \Big(3r^5 +r^4+12r^3-16r^2
-4(r^2+3r+2)r^2\ln r +r-1 \Big),\\
 {\sigma}_{L}^{\COcPz}
=&
-F  \frac{1+r}{(1-r)^4}\Big(r^4
-22r^3+16r^2
+8(r+2)r^2\ln r +6r-1 \Big)\,,
\end{align}
\end{subequations}
with $F = \frac{32 \pi ^2 \alpha^2 \alpha _s  e_c^2   }{3
	s^2 m_c^3} \langle
  \mathcal{O}^{J/\psi}(\COcPz)  \rangle $. We notice that when $r\to 0$, ${\sigma}_{T}^{\COcPz} \to 2 {\sigma}_{L}^{\COcPz}$, and in this limit the $\chi$ dependence tends to drop out.

\subsection{Quarkonium Energy Correlator in $pp$ collisions}
Similar measurements can be carried out in $pp$ collisions. For illustration, we simplify our analysis by assuming that the $J/\psi$ is predominantly produced through the gluon fragmentation
\begin{align}\label{eq:pqcdfac}
d\sigma_{A+B\to J/\psi+X}(p_\psi) \approx &
\int_0^1 dz
d{\hat{\sigma}}_{A+B\to g+X}({\hat p}/z,\mu_F)
 D_{g\to J/\psi}(z,\mu_F)
\,,
\end{align}
where $d\hat\sigma_{A+B\to g+X}$ is the partonic production cross section for hadrons
$A$ and $B$ to produce parton $g$, and $p_\psi$ is the momentum of $J/\psi$. We introduced the notation $\hat p^\mu=(p_\psi^+,0,0_\perp)$ and $z$ is the $J/\psi$ momentum fraction. In NRQCD, the fragmentation function $D$ can be further factorized as
\begin{align}
D_{g\to J/\psi}(z,\mu_F) = &
\sum_{n}
d_{g\to c{\bar c}[n]}(z,\mu_F) \langle  \mathcal{O}^{J/\psi}(n)  \rangle
\,,
\end{align}
where $d_{g\to c{\bar c}[n]}$ can be calculated perturbatively.
The dominant channels are $n=\COaSz$, $\COcSa$, $\COcPj$, where the $n=\COaSz$ and
$\COcPj$ contribute leadingly in $\alpha_s$ to the $\Sigma_{P.T.}(\cos\chi)$ by emitting one hard gluon, while the soft radiation in the $\COcSa$ channel constitutes the major part of the $\Sigma_{N.P.}(\cos\chi)$.

At LO in $\alpha_s$, one hard gluon is emitted in the $\COaSz$ and $\COcPj$ channels and we have
\begin{align}
d^{\mathrm{LO}}_{g\to c\bar c[n]}(z,\mu_F) &=\int ds \hat d_{g\to c\bar c[n]}(z,s,\mu_F) \,,
\end{align}
where $s=(p_\psi+k)^2$ and $k$ denotes the momentum of the hard gluon.
Then the $J/\psi$ energy correlator due to the hard gluon reads
\begin{align}
\Sigma_{P.T.}(\cos{\chi})=& \frac{1}{M}\sum_{n} \int_0^1 dz
d{\hat{\sigma}}_{A+B\to g+X}({\hat p}/z,\mu_F) \int ds \frac{k\cdot p_\psi}{2m_c}
\nonumber\\
&\hspace{-1cm} \times
\hat d_{g\to c\bar c[n]}(z,s,\mu_F) \delta\Big(\cos{\chi}-\frac{2\sqrt{2}m_c k^+}{k\cdot p_\psi}+1\Big)
 \langle  \mathcal{O}^{J\psi}(n)  \rangle.
\end{align}
Here we focus on the region in which $\cos\chi >0$ and therefore we ignore the contribution from a parton in $\hat{\sigma}_{A+B\to g+X}$ to hit the detector. A complete calculation will be presented in a future work.
 Now,
using $s=(p_\psi+k)^2$ and $k^+=p_\psi^+ (1-z)/z$, we can rewrite $\Sigma_{P.T.}(\cos{\chi})$ as
\begin{align}\label{eq:pp}
\Sigma_{P.T.}(\cos{\chi}) =& \sum_{n} \int_0^1 dz
d{\hat{\sigma}}_{A+B\to g+X}({\hat p}/z,\mu_F)
 \hat D_{g\to c\bar c[n]}(z,\cos{\chi},\mu_F)
\langle  \mathcal{O}^{J/\psi}(n)  \rangle,
\end{align}
with
\begin{align}
\hat D_{g\to c\bar c[n]}(z,\cos{\chi},\mu_F) =   \frac{2}{M}\int d(k\cdot p_\psi)
 \Big(\frac{k\cdot p_\psi}{2m_c}\Big)^3 \frac{z}{1-z}
 \hat d_{g\to c\bar c[n]}(z,k\cdot p_\psi,\mu_F)
 \delta\Big(k\cdot p_\psi-\frac{4m_c^2 }{1+\cos{\chi}}\frac{1-z}{z}\Big) .
\end{align}
For $n=\COaSz$ and $\COcPj$, we find
\begin{align}
\hat D_{g\to c\bar c[\COaSz]}(z,y,\mu_F) =&  \frac{5 \alpha_s^2}{16} \frac{1}{6m_c^3}
\frac{(1-z)^2}{ (y+1) ((y-1) z+2)^2}
 \times
 \Big( (y-1)^2 z^2+2 (y-1) z+2 \Big), \\
\hat D_{g\to c\bar c[\COcPj]}(z,y,\mu_F) =&  \frac{5 \alpha_s^2}{16} \frac{1}{6m_c^5}
 \frac{ 1 }{(y+1)  ((y-1) z+2)^2}
 \times
\Big(-2 \left(y^3-10 y+9\right) z^3
 +\left(3 y^2-14 y+25\right) z^2
 \nonumber\\
&\hspace{.5cm}
+(y-1)^2 (2 y+5) z^4
 +2 (5 y-7) z+6\Big),
\end{align}
with $y=\cos\chi$. Inserting the above results into Eq.~\eqref{eq:pp}, we obtain the $\Sigma_{P.T.}$.

For $\COcSa$, we write
\begin{align}\label{eq:pp-np}
\Sigma_{N.P.}(\cos{\chi}) =&  \int_0^1 dz
d{\hat{\sigma}}_{A+B\to g+X}({\hat p}/z,\mu_F)
 \hat D_{g\to c\bar c[\COcSa]}(z,\cos{\chi},\mu_F)
\langle  \mathcal{O}^{J/\psi}(\COcSa)  \rangle,
\end{align}
where
\begin{align}\label{eq:D3S18}
\hat D_{g\to c\bar c[\COcSa]}(z,\cos{\chi},\mu_F) =& \int {d \Phi_{gg} }\frac{2k_1^0}{M}
\delta(\cos{\chi} - \cos{\theta})
\sum_{\lambda}
d^{\mathrm{LO}}_{g\to c\bar c[^3S_{1,\lambda}^{[8]}]}(z,\mu_F) \,
\frac{A^{\COcSa}_\lambda(\theta,\phi)}{\langle  \mathcal{O}^{J/\psi}(\COcSa) \rangle},
\end{align}
here $d \Phi_{gg}$ is the phase space for two final state gluons emitted. $A^{\COcSa}_\lambda$ denotes the transition amplitude for $c{\bar c}[^3S_{1,\lambda}^{[8]}] \to J/\psi + gg$, and $\theta$ and $\phi$ are the polar and
azimuthal angles of one of the emitted soft gluons in the $J/\psi$ helicity frame. $k_1$ and $k_2$ represent the momentum of the gluons.

At the leading order in
$v$, the $c\bar c[^3S_{1,\lambda}^{[8]}]$ pair will transmit into the $J/\psi$ via two $E_1$ gluon emissions. Using the heavy quark spin symmetry, we
find $A^{\COcSa}_\lambda(\theta,\phi)$ is also independent of $\theta$ and $\phi$. Therefore, we have
\begin{align}\label{eq:G2}
A_{\lambda}^{\COcSa}(\theta,\phi) = \frac{1}{4\pi} \frac{1}{3} G_2,
\end{align}
where $G_2$ is a solid angle independent function and satisfies $\frac{1}{4\pi}\int d\Phi_{gg} G_2 \sim {\cal O}(\langle {\cal O}^{J/\psi}(\COcSa) \rangle)$. Inserting Eq.~\eqref{eq:G2} into Eq.~\eqref{eq:D3S18}, we find
\begin{align}\label{eq:D3S18-2}
\hat D_{g\to c\bar c[\COcSa]}(z,\cos{\chi},\mu_F) =& \int {d \Phi_{gg} }\frac{G_2}{4\pi}  \frac{2k_1^0}{M}
\delta(\cos{\chi} - \cos{\theta})
d^{\mathrm{LO}}_{g\to c\bar c[\COcSa]}(z,\mu_F) \,
\frac{1}{\langle  \mathcal{O}^{J/\psi}(\COcSa) \rangle} \nonumber\\
\equiv& \bar v_2 \times  \frac{1}{2} d^{\mathrm{LO}}_{g\to c\bar c[\COcSa]}(z,\mu_F),
\end{align}
here $m_c {\bar v}_2$ is the average energy of one of the emitted soft gluons. And
\begin{align}
d^{\mathrm{LO}}_{g\to c\bar c[\COcSa]}(z,\mu_F)=\frac{\pi \alpha_s}{24m_c^3}\delta(1-z).
\end{align}
Now using Eqs. \eqref{eq:pp-np} and \eqref{eq:D3S18-2}, we get the $\Sigma_{N.P.}$.

\end{widetext}

\end{document}

%% file: paper.bbl
\providecommand{\href}[2]{#2}\begingroup\raggedright\endgroup